\documentclass[letterpaper, 10 pt, conference]{IEEEtran}
\IEEEoverridecommandlockouts
\usepackage{cite}
\usepackage{amsmath,amssymb,amsfonts}
\usepackage{algorithmic}
\usepackage{graphicx}
\usepackage{textcomp}
\usepackage{tcolorbox}
\usepackage{epsfig}
\usepackage{bmpsize}
\usepackage{xcolor}
\usepackage{lipsum}
\def\BibTeX{{\rm B\kern-.05em{\sc i\kern-.025em b}\kern-.08em
    T\kern-.1667em\lower.7ex\hbox{E}\kern-.125emX}}
\usepackage{filecontents}
\usepackage{times}
\usepackage{balance}
\usepackage{authblk}
\usepackage{epsfig}  
\usepackage[normalem]{ulem}
\usepackage{amscd}
\usepackage{microtype}
\usepackage{array}
\usepackage{chngpage}
\usepackage{graphics}
\usepackage{epsfig}
\usepackage{multirow}
\usepackage{graphicx}
\usepackage{caption}
\usepackage{listings}
\usepackage{subfigure}
\usepackage{adjustbox}
\usepackage{tabularx}
\usepackage{epstopdf}
\usepackage{threeparttable}
\usepackage{ulem}
\usepackage{lipsum}
\usepackage{float}
\usepackage{pifont}
\usepackage{bm}
\usepackage{booktabs}
\usepackage{color,xcolor}
\usepackage{verbatim}
\usepackage{amssymb}
\usepackage{pifont}
\usepackage{array,multirow}
\usepackage{makecell}
\newcommand{\xmark}{\ding{55}}%
\usepackage{tikz}
\usepackage{CJK}
\usepackage{float}

\newcounter{rownumbers}

\newcommand{\ignore}[1]{}

\begin{document}

\title{ \bf  SoK: A Modularized Approach to Study the Security of Automatic Speech Recognition Systems}

\author[1,2]{\normalsize Yuxuan Chen}
\author[3,4]{Jiangshan Zhang}
\author[3,4]{Xuejing Yuan}
\author[5]{Shengzhi Zhang} 
\author[3,4]{Kai Chen}
\author[6]{XiaoFeng Wang}
\author[1,2]{Shanqing Guo}
\affil[1]{\normalsize{Key Laboratory of Cryptologic Technology and Information Security, Ministry of Education, Shandong University, China}}
\affil[2]{\normalsize{School of Cyber Science and Technology, Shandong University, China}}
\affil[3]{\normalsize{SKLOIS, Institute of Information Engineering, Chinese Academy of Sciences, China}}
\affil[4]{School of Cyber Security, University of Chinese Academy of Sciences, China}
\affil[5]{Department of Computer Science, Metropolitan College, Boston University, USA}
\affil[6]{School of Informatics, Computing and Engineering, Indiana University Bloomington, USA}

\maketitle
\thispagestyle{plain}
\pagestyle{plain}

\begin{abstract}
With the wide use of Automatic Speech Recognition (ASR) in applications such as human machine interaction, simultaneous interpretation, audio transcription, etc., its security protection becomes increasingly important.  Although recent studies have brought to light the weaknesses of popular ASR systems that enable out-of-band signal attack, adversarial attack, etc., and further proposed various remedies (signal smoothing, adversarial training, etc.), a systematic understanding of ASR security (both attacks and defenses) is still missing, especially on how realistic such threats are and how general existing protection could be. In this paper, we present our systematization of knowledge for ASR security and provide a comprehensive taxonomy for existing work based on a modularized workflow. More importantly, we align the research in this domain with that on security in Image Recognition System (IRS), which has been extensively studied, using the domain knowledge in the latter to help understand where we stand in the former. Generally, both IRS and ASR are perceptual systems. Their similarities allow us to systematically study existing literature in ASR security based on the spectrum of attacks and defense solutions proposed for IRS, and pinpoint the directions of more advanced attacks and the directions potentially leading to more effective protection in ASR. In contrast, their differences, especially the complexity of ASR compared with IRS, help us learn unique challenges and opportunities in ASR security. Particularly, our experimental study shows that transfer attacks across ASR models is feasible, even in the absence of knowledge about models (even their types) and training data.

\end{abstract}

\section{Introduction}
Automatic Speech Recognition (ASR) systems are becoming increasingly popular these days due to the convenience they provide to the users to manage smart devices and interact with cloud services. Today, intelligent voice control (IVC) devices enabled with voice recognition skills like Google Home, Amazon Echo, Apple HomePod are already deeply involved in our daily life, helping unlock the doors of houses or cars, make online purchase, send messages, and etc. The availability of ASR services such as Google Cloud Speech-to-Text, Amazon Transcribe, Microsoft Bing Speech Service and IBM Speech to Text further allows their users to conveniently integrate their APIs to control smart devices, conduct audio transcription, text analysis, video analysis and etc.

In the meantime, ASR systems today are exposed to various security risks. Previous researches~\cite{38,39,47,48} show that the hardware for pre-processing input signals for the ASR system can be exploited to stealthily command the systems: for example, Sugawara et al.~\cite{48} demonstrates the injection of malicious commands to ASR through specially manipulated laser light by exploiting the photo-acoustic effect. Further researches~\cite{7,18,19,29} show that the way the ASR system extracts feature from audio signals could be manipulated: Carlini et al.~\cite{7} invert the MFCC~\cite{200} features of malicious commands back to audio and then cover the commands with noise to get the attack audio samples uninterpreted to human beings but recognizable to the ASR system and the IVC device (such as Google Speech API and Google Assistant on smartphone). Recent effort~\cite{6} has been made to find adversarial examples from ASR systems and IVC devices, which have been extensively studied in image recognition systems~\cite{109,116,118,121,128,130}. Specifically, by exploiting the inherent vulnerability in the machine learning algorithm, one can generate an adversarial example with unnoticeable perturbations but causing misclassification on the target neural network model. For example, Yuan et al.~\cite{4} generate audio adversarial samples to attack the state-of-the-art Kaldi ASR system~\cite{kaldi}, which are shown to be less distinguishable to human. Such attacks could have serious consequences, potentially putting the victims in a grave danger: e.g., the attackers compromising home IVC devices could open the garage door without victim's awareness.

Like ASR, image recognition is another popular system powered by machine learning and therefore also subject to the threat of various attacks, and some of them were exploiting sensor vulnerability~\cite{136,137} with most of them were adversarial attacks~\cite{109,116,118,121,128,130}. Actually, adversarial attacks have already been extensively investigated in the image domain, while much fewer studies have been done on the audio side, mainly due to the unique complexity of audio signals and ASR systems. Intuitively, both the ASR system and the image recognition system (IRS) are \textit{perceptual system}, with similar functionalities of perceiving, processing and recognizing. So one may ask \textit{whether such similarities present a unique perspective for better understanding ASR security and gaining insights into the research directions that could lead to more effective protection.}  Up to our knowledge, such a cross-domain, systematic analysis for systematization of knowledge in ASR security has not been done before. 


In this paper, we report our effort to systematically analyze the security threats to ASR systems. We first present our modularized system workflow for ASR system. Then we describe a comprehensive taxonomy, based upon the modularized workflow, for known attacks on ASR and existing protection. After that, we compare and analyze across existing attacks and defenses in both image and audio domains for each component under our modularized model. This analysis has brought to light new insights about challenges and opportunities in research on ASR security. For example, the idea of image down-scale attack may be borrowed to conduct effective audio adversarial attack (See Section~\ref{sec:comparewithimageattack}). Particularly, our study shows that given the complexity of ASR, transfer attacks across multiple, block-box systems are still feasible, as evidenced by our experimental results in Section~\ref{sec:transfer}.  

\vspace{2pt}\noindent\textbf{Comparison with related research}. Concurrently but independently, Hadi et al. have also reported their SoK work on ASR Security~\cite{126}, which has come to our attention recently. However, their study significantly differs from ours in the following aspects. First, departing from a key conclusion drawn by the related work~\cite{126}, which finds transfer attacks in ASR to ``almost universally fail'', our research demonstrates that a transferable adversarial attack across different ASR systems (e.g., from DeepSpeech \cite{DeepSpeech} to the commercial black-box models like Microsoft Bing ASR~\cite{Microsoftazure}) is entirely feasible under similar settings~\cite{126}, using the gradient-based approach~\cite{6}. We found that the difficulty of transferability is likely caused by the weak features of the target phrases in the generated adversarial examples, especially in the digital space, rather than the inherent properties of ASR as stated in the related work~\cite{126}. As evidence, we have released the adversarial examples used in our experiments and several demos showing the successful transfer across ASR models (https://github.com/AsrAttackSok/asr-attack-sok). Second, we present a unique perspective that compares with the adversarial learning on image recognition to enable in-depth understanding of ASR security and envision of potential future directions, as summarized through a set of key take-aways (see Section~\ref{sec:comparewithimageattack} and~\ref{sec:comparewithimagedefense}); such insights are missing in the related study~\cite{126}, even though it also draws parallel between audio and image-based adversarial learning. Finally, we formulate the findings of existing studies along the modularized workflow (see Section~\ref{sec:attack} and Section~\ref{sec:defense}), which helps clearly identify the part of the problem space that has already been intensively investigated and further the areas that need more attention; by comparison, the related work~\cite{126} systematizes the existing research based on the threat model, thereby failing to offer such an insight.

\vspace{2pt}\noindent\textbf{Contributions}. The contributions of this paper are as follows:

\vspace{1pt}\noindent$\bullet$\textit{ Modularized taxonomy of existing studies}. We present a systematic and comprehensive taxonomy on existing attacks and defense on ASR security using a modularized workflow, which helps better understand the focuses and findings of such existing studies, and further sheds light on new directions towards more advanced attacks and better protection.

\vspace{1pt}\noindent$\bullet$\textit{ A comparison study on ASR and IRS security}. We offer a new perspective to look at ASR security through the lens of well-studied IRS security. Such a comparison results in take-aways that pinpoint the directions leading to more optimized attacks and more effective defense on audio systems.

\vspace{1pt}\noindent$\bullet$\textit{ New understanding of transferable audio adversarial attack}. We demonstrate that using gradient based approach~\cite{6}, \textit{transferable} audio adversarial attack is feasible even on black-box ASR systems with unknown training sets, parameters and inside architectures. Based on our experimental results, we provide our understanding of the difficulty in transferable adversarial attacks on ASR, and the ways to make such attack more likely to succeed.

\section{Methodology}
\label{sec:methodology}
To better facilitate the systematic study of existing literature for attacks and defense mechanisms towards ASR, we first propose our modularized workflow to model current ASR system and summarize the properties in each component for detailed systematization. Finally in this section, we discuss the scope of this paper.

\subsection{Modeling ASR}

\begin{figure*}
\centering
\epsfig{figure=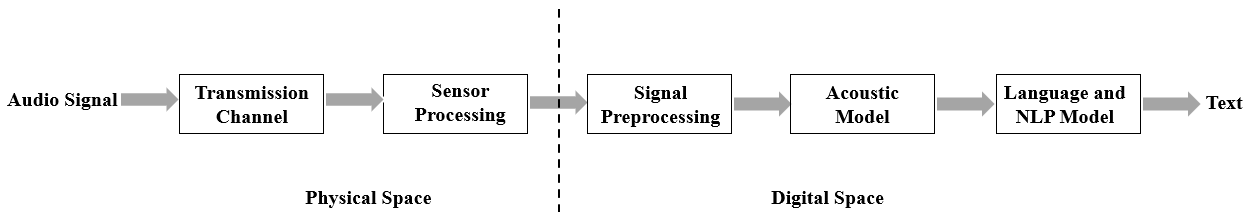, width=1\textwidth} 
\caption{A Modularized ASR System Workflow}
\label{fig:modular}
\vspace{-0.2cm}
\end{figure*}

In the physical world, ASR systems typically begin by extracting desired features from the physical input, e.g., acoustic signal, and utilize pre-trained neural network models to decode/interpret the extracted features. To systematically investigate existing attacks and defenses in ASR, we consider a physical scenario, i.e., the input signal transmitted over-the-air and then received by the corresponding device for processing. The components considered in this systematization are summarized as show in Figure \ref{fig:modular}\footnote{In some scenarios, one or several components may be missing for a particular ASR system. For instance, an online ASR system (offering decoding service) does not include the transmission channel or sensor processing.}.

\vspace{2pt}\noindent\textbf{Transmission Channel.} Before a physical audio signal from the source reaches the device end, it is first transmitted over-the-air in an open channel, and the surroundings may impact the audio signal transmission in various ways. First, the strength of acoustic signal gets attenuated during transmission in a medium as distance increases, thus imposing the constraint on the effective distance for the device to capture sufficient energy from the signal. Second, the acoustic signal reflects against surrounding objects during transmission, so the actual signal captured by the device is always a composition of signals from various sources, rather than the pure original signal. Finally, the ambient noise in the open air environment will also interference the original signal during the transmission, which potentially destruct the speech information contained in the signal and lead to misunderstanding by ASR systems. As a consequence, the module of transmission channel is a major concern when conducting attacks against ASR in the physical space, which will be further discussed in Section~\ref{sec:transmission with image}.

\vspace{2pt}\noindent\textbf{Sensor Processing.} Sensors are devices that take audio signal from the surroundings as input, and convert it into digital representation for processing. In particular, sensors for ASR system, like microphone, etc., rely on different electrical transducers to capture acoustic signal, and convert it into analog electrical signal. Typically, all transducers introduce random noise, which then needs to be reduced by filters. Meanwhile, the useful information inside the converted electrical signal needs to be reinforced by amplifiers. Finally, the analog-to-digital converter (ADC) transforms the filtered/amplified electrical signal into a digital signal by sampling and quantization.

\vspace{2pt}\noindent\textbf{Pre-processing.} A series of pre-processing operations then are necessary with the general purposes of removing irrelevant audio information and emphasizing the useful speech features in the digital signal. Generally, ASR system needs to apply denoising to the digital signal to filter out various noises introduced from surroundings or electronic components\footnote{The filtering here works in the original domain, i.e., time domain for the acoustic signal.}. For instance, adaptive Least Mean Square filter \cite{201} or Wiener filter \cite{202} can be used to remove environmental noises from the acoustic signal. Moreover, transformation over the digital audio signal is also necessary. For example, voice activity detection (VAD) identifies segments with voice (by filtering out silent segments) inside the acoustic signal, and further divides those segments into overlapping frames. 

Furthermore, ASR system needs to adjust sampling rate, either downsampling or upsampling, to reduce or enrich the information inside the digital signal. Finally, the useful information in the digital signal should be well enhanced, typically done in the frequency domain, by pre-emphasizing the energy at the high frequency to balance the spectral tilt (less energy at the higher frequency than that at the lower frequency for vowels). Fast Fourier Transformation (FFT) is also conducted on each frame of the acoustic signal to compute the frequency spectrum, where acoustic features are further extracted using either filter bank or MFCC (Mel-Frequency Cepstral Coefficients). Both of them extract magnitude\footnote{Human perception is less sensitive to phase information.} from the frequency spectrum at the mel scale and calculate the log of the square of the magnitude. Then MFCC applies Discrete Cosine Transform (DCT) to decorrelate the coefficients of the above logarithmic results.

\vspace{2pt}\noindent\textbf{Acoustic Model.} The acoustic model computes the statistical relationship between the acoustic features from the above and the acoustic units, such as phoneme or character. Transitionally, Gaussian Mixture Model (GMM) was popularly used in acoustic model. Nowadays, neural networks have been widely used in acoustic model, causing decorrelation (by MFCC) unnecessary due to their capability of handling even highly correlated inputs. Hence, filter bank approach to extract acoustic features is becoming dominant. 

\vspace{2pt}\noindent\textbf{Language Model.} Language model builds phonemes into words by computing the probability of sequences of words, and uses context information to distinguish words with similar pronunciation. Traditionally, statistical language model was used like N-grams and Hidden Markov Model (HMM)\footnote{Since GMM-HMM is highly susceptible to correlated input, MFCC was co-evolved together due to its decorrelation.}, but nowadays neural language models, e.g., based on Recurrent Neural Network (RNN), become overwhelming due to their effectiveness. 

\vspace{2pt}\noindent\textbf{NLP Model.} Natural Language Processing (NLP) is for machines to ``understand'' the decoded text from language model and trigger the corresponding actions if necessary. Given the tasks involved in the voice-controlled devices are intermediate level difficulty, part-of-speech tagging and dependency parsing are commonly used techniques to derive the semantics from the decoded text. The former approach tags words with similar grammatical properties, while the latter builds the parse tree out of the decoded text using either dependency parsing (relationships among words) or constituency parsing (probabilistic context-free grammar).

\subsection{Attack Properties}
\label{sec:attackproperties}
The fundamental goal of attacking ASR is to control it in an attacker-desired fashion. Simultaneously, such attack should be as stealthy as possible, better to be unnoticeable to people nearby. For instance, the attackers may craft a piece of audio signal, that when played, causes the users unable to wake up the voice controlled devices. The attacker can also manipulate the raw input, thus inducing the ASR into generating different results than human perception. We consider the below aspects when evaluating existing attacks against ASR. 

\subsubsection{Attack Vectors}
Generally, the attack against ASR involves manipulation of the raw signals in a particular way, e.g., injecting out-of-band signal decoded as the targeted text, adding crafted perturbations as adversarial examples, etc. To design such manipulation, one or several specific modules of ASR need to be investigated as the main ``target'' to process such manipulation in the attacker-desired way, e.g., the sensor processing module for the former and the acoustic model for the latter. Hence, in this paper, we elaborate the attack vectors against ASR using the above insight, that is, the main target module where the attacks exploit, which aligns well with our modularized approach to study ASR. 

\vspace{2pt}\noindent\textbf{Transmission channel.} Since the input of the sensor processing module is always a mixed input of acoustic signals, it is possible for attackers to broadcast their hacking audio overlaying with the human voice, thus causing the ASR compromised. Furthermore, to attack ASR physically, the transmission channel needs to be considered together with the main target module to address the distortion incurred by over-the-air transmission. The distortion incurred by the speaker used to generate the hacking audio and the sensor used to receive the acoustic signal also needs to be considered.

\vspace{2pt}\noindent\textbf{Sensor processing.} Attackers may leverage out-of-band signal, e.g., ultrasound, laser, electromagnetic wave, etc., other than sound, to trigger the imperfect components inside sensors. The transduction attack \cite{203} can happen when the sensors' output (manipulated by attackers via out-of-band signal) is decoded as desired by attackers.

\vspace{2pt}\noindent\textbf{Pre-processing.} Loss exists when acoustic features are extracted from the raw audio signal during pre-processing. Hence, it is highly possible that multiple pieces of audio signals correspond to the same or similar acoustic features used by acoustic model to decode. With the knowledge of the acoustic features of the desired commands, the attackers may ``reversely'' identify one piece of raw signal among many, which still preserves most of the acoustic features as the desired commands, but uninterpretable to people any more. Besides feature extraction, voice activity detection (VAD) utilizes a threshold to determine the existence of activity in audio segments, which may help attackers to determine the optimal energy of their attacks to go through. Meanwhile, pre-emphasis on high frequency signal can be leveraged by attackers to amplify their injected signal as they desire, thus ``helping'' their attacks to succeed.

\vspace{2pt}\noindent\textbf{Acoustic model.} The neural networks are known to be vulnerable to adversarial examples \cite{118}, which cause difference between human perception and machine recognition. Targeting at the neural networks in acoustic model, the attackers can craft adversarial examples to mislead ASR into recognizing their desired results, but at the same time uninterpretable (or interpreted differently than the attackers-desired results) to people.

\vspace{2pt}\noindent\textbf{Language model and/or NLP model.} For ASR with language model and NLP model, attacks could be designed to explore words with similar pronunciation or phrases with similar semantics. The goal for such attacks typically is not to cause different perception between human and machine, but to fool the NLP model to mis-trigger back-end operations.

\subsubsection{Attack Metrics} A suit of metrics could be used to evaluate the attacks against ASR. Below we choose the ones frequently used in literature to characterize existing attacks. 

\vspace{2pt}\noindent\textbf{Consequence.} The attacks can simply cause ASR to mis-decode the audio into any wrong text (untargeted attack) or intentionally mislead ASR to decode the audio into the desired text (targeted attack).

\vspace{2pt}\noindent\textbf{Knowledge.} The attacker may have full knowledge (e.g., details of each module in ASR), partial knowledge (e.g., one or several, but not all modules), or no knowledge of ASR, corresponding to white box, gray box, and black box attacks, respectively.

\vspace{2pt}\noindent\textbf{Domain.} The attacks can begin by feeding ``digital'' input into the pre-processing module or broadcasting ``physical'' signal over-the-air into the sensor processing module, referred to as digital attack and physical attack, respectively.



\vspace{2pt}\noindent\textbf{Generality.} Attacks against ASR might be universal, which means the same perturbation could be added on any given audio sample to make ASR misclassify. In contrast, specific attack means that the perturbation can only be added to a specific audio signal to fool ASR.

\vspace{2pt}\noindent\textbf{Human perception study.} A straightforward way to evaluate the stealthiness of the attacks is to conduct survey\footnote{Typically IRB is required for such survey.}, asking participants' perception on the attacks to evaluate the stealthiness. 

\vspace{2pt}\noindent\textbf{SNR.} An alternative used to evaluate the stealthiness of attacks is to measure the Signal to Noise Ratio (SNR) of the attack. However, SNR only reflects the revision (either the max or average) over the original signal, rather than the human perception on the actual attack audio. 

\subsection{Defense Properties} Currently, there are limited number of defense solutions for attacks against ASR, including both detection and prevention. In our systematization, we consider the below aspects when studying those existing defense solutions. 

\subsubsection{Defense methods} Considering the attacks against ASR might mainly target a particular module, defense can be done by enhancing the dependability of the module or adding an extra detection/prevention layer ahead of the module. Another spectrum of defense explores the idea of N-version \cite{75}, by diversifying either the input or modules and comparing the output to detect anomaly.

\subsubsection{Defense metrics} Below we discuss the metrics used in this paper to evaluate existing defense solutions.

\vspace{2pt}\noindent\textbf{Generality.} The defense solutions can be generic to attacks targeting various modules or specific to attacks targeting a particular module of ASR. For instance, guarding sensor processing module is specific to attacks targeting on sensor processing module, but may not work against attacks targeting other modules.

\vspace{2pt}\noindent\textbf{Knowledge.} The defense solutions can be designed with the assumption of the full knowledge of ASR (e.g., to enhance an existing module), partial knowledge (e.g., to integrate an extra module inside ASR), or none knowledge (e.g., to add an extra module ahead of ASR). Similar as attacks, they are referred to as white box, gray box, and black box defense respectively.

\subsection{Scope of this SoK}
Our systematization focuses on those attacks that directly target/explore vulnerabilities in ASR systems, and the corresponding defense solutions against such attacks. Although there exists a large amount of works closely related to ASR, they are out of the scope of this paper. First, from the acoustic signal, speaker verification system extracts features closely related to the identity of speakers, rather than those related to semantics. Since it performs different task than speech recognition system, in our systematization, we do not consider those attacks or defense works in the speaker verification domain \cite{95,96,97,98,99,100,101,102,103,104}. Second, we do not consider exploiting vulnerabilities in other applications, operating systems, or even hardware \cite{88,90,91,92,93,94} to indirectly attack ASR, since our intention is to study the inherent ``vulnerabilities'' residing in ASR itself. Similarly, those works leveraging acoustics signals to attack other software, system, or platforms \cite{81,82,83,84,85,86,87} are also out of scope of our paper. Finally, we do not consider the privacy leakage attacks associated with audio signal, e.g., utilizing the echo reflected from nearby objects or faces to pinpoint the surroundings or face of the victims \cite{105,106,107,108}.

\section{Systematization of Attacks}
\label{sec:attack}
In this section, we present the systematization of existing attacks against ASR based on our modularized approach, i.e., according to the modules these attacks exploited. In Table~\ref{Attack}, we elaborate the literature investigated in this paper based on the attack properties discussed in Section \ref{sec:attackproperties}. 

\begin{table*}[ht]
\centering
\small
\caption{Attack Taxonomy.}
\label{Attack}
\begin{tabular}{m{2.3cm}<{\centering}|m{2.5cm}<{\centering}|m{1.7cm}<{\centering}|m{1.5cm}<{\centering}|m{1cm}<{\centering}|m{1.3cm}<{\centering}|m{1.3cm}<{\centering}|m{2.5cm}<{\centering}}
\Xhline{3\arrayrulewidth}
\textbf{Attack} & \textbf{Target} &\textbf{Consequence}  & \textbf{Knowledge} & \textbf{Domain} & \textbf{Generality} & \textbf{Perception Survey} & \textbf{SNR(dB)}  \\ 
\hline
\cite{38,39,40,41,42,43,44,45,46,47,48} & Sensor  & Targeted & Black-box & Physical & Specific & Not applied & Not applied   \\ 
\hline
\cite{10} & Transmission & Targeted  & White-box & Physical &  Specific &  \xmark  & -2.9$\thicksim$17.3 \\ 
\hline
\cite{14} & Transmission  & Targeted & White-box & Physical & Specific & \checkmark & 0.2$\thicksim$14.6   \\ 
\hline
\cite{21} & Transmission & Targeted & White-box & Physical & Specific & \checkmark & ?   \\ 
\hline
\cite{27} & Transmission & Untargeted & Black-box & Physical & Specific & \xmark &    ?  \\ 
\hline
\cite{18} & Preprocessing & Targeted & Black-box & Physical & Specific & \xmark &  ?  \\ 
 \hline
\cite{19} & Preprocessing &Targeted  & Black-box & Digital & Specific &\checkmark  & ?   \\ 
 \hline
\cite{29} & Preprocessing & Untargeted & Black-box & Physical & Specific & \checkmark & ?   \\ 
 \hline
\cite{1} & Acoustic & Targeted & Black-box  & Digital & Specific &  \checkmark & ? \\ 
\hline
\cite{2} & Acoustic & Targeted & Black-box & Digital & Specific & \checkmark  &  14.0$\thicksim$22.0 \\ 
\hline
\cite{3} & Acoustic & Targeted & White-box & Digital & Specific & \checkmark & 15.88$\thicksim$21.76     \\ 
\hline
\cite{4} & Acoustic & Targeted & White-box & Physical & Specific & \checkmark & 14$\thicksim$18.6(digital) 1.3$\thicksim$1.5(physical)  \\ 
\hline
\cite{5} & Acoustic & Targeted & White-box & Digital & Specific & \checkmark & ?  \\ 
\hline
\cite{6}& Acoustic & Targeted & White-box & Digital & Specific & \xmark & -15$\thicksim$-45    \\ 
\hline
\cite{7} & Acoustic & Targeted & White-box & digital & Specific & \checkmark & ?  \\ 
\hline
\cite{8}& Acoustic & Untargeted & White-box & digital & Specific & \checkmark & ?      \\ 
\hline
\cite{9} & Acoustic & UT/T & White-box & digital & Universal & \xmark & 18.489(UT) 19.69(T)     \\ 
\hline
\cite{11} & Acoustic  & Targeted & White-box & physical & Specific & \xmark & 24.1$\thicksim$27.9   \\
\hline
\cite{12} & Acoustic  & Targeted & White-box & physical & Specific & \checkmark & ?   \\ 
\hline
\cite{13} & Acoustic  & Targeted & Black-box & physical & Specific & \checkmark & 9.39$\thicksim$13.36(digital)
7.86$\thicksim$12.10 (physical)  \\ 
\hline
\cite{15} & Acoustic  & Targeted & White-box & digital & Specific & \xmark & ?   \\ 
\hline
\cite{16} & Acoustic  & Targeted & Black-box & digital & Specific & \xmark & ?   \\ 
\hline
\cite{17} & Acoustic  & Targeted & White-box & digital & Universal & \xmark & -32(mean)  \\ 
\hline
\cite{20} & Acoustic & Targeted & White-box & digital & Specific & \xmark &  31.9(mean)  \\ 
 \hline
\cite{22} & Acoustic & Targeted & Black-box & digital & Specific & \xmark & ?   \\ 
 \hline
\cite{28} & Acoustic & Untargeted & White-box & Physical & Universal & \xmark &  -1.02$\thicksim$36.64  \\ 
  \hline
\cite{30} & Acoustic & Targeted &White-box  & Physical & Specific & \xmark &  24.488(mean)  \\ 
   \hline
\cite{31} & Acoustic & Targeted & White-box & Digital & Specific & \xmark &  -28.51(mean)  \\ 
   \hline
\cite{32} & Acoustic & Untargeted & White-box & Digital & Specific & \xmark & ?   \\ 
    \hline
\cite{34} & Acoustic  &Untargeted  & White-box & Digital & Universal & \xmark& ?   \\ 
    \hline
\cite{35} & Acoustic & Targeted & White-box & Digital & Universal & \xmark &?    \\ 
    \hline
\cite{37} & Acoustic & Untargeted &  Black-box & Digital & Specific  & \checkmark & ?   \\ 
     \hline
\cite{23,25,26} & Language & Targeted & Black-box & Digital & Specific & \checkmark &  ?  \\ 
 \hline
\cite{24} & Language & Targeted & Black-box & Digital & Specific & \xmark &  ?  \\
  \hline
\Xhline{3\arrayrulewidth}
\end{tabular}
\end{table*}

\subsection{Targeting Transmission Channel}

As discussed in Section \ref{sec:methodology}, transmission channel is a major concern when conducting physical attacks against ASR systems. During the past few years, many related works~\cite{4,7,10,11,12,13,14,21,27,28,29,30} have emerged to enhance the robustness of the crafted audio adversarial examples in the physical space by exploiting transmission channel. 

In \cite{10,11,27,30}, the authors apply Room Impulse Response (RIR) to simulate the transmission of an acoustic signal from the loudspeaker to the microphone, to enhance the physical robustness of the adversarial examples. In particular, the signal transmission is modeled as a convolution of the original audio signal with the room-dependent RIR. The authors in \cite{14} discuss three approaches, including a band-pass filter, room impulse response, and white Gaussian noise, to generate a robust adversarial example that can attack DeepSpeech over-the-air.

In \cite{21}, the authors first obtain an initial perturbation
that captures the impact from the core distortion based on only a small set of prior measurements, and then take advantage of a domain
adaptation algorithm to refine the perturbation to further improve the attack distance and reliability. In \cite{4,13,30}, the authors introduce white Gaussian noise into the adversarial example generation process to improve the physical robustness. They believe that the white Gaussian noise can well mimic the circumstances of some random processes that appear in the transmission channel. Therefore, they do not need to build a specific noise model, which could be too device-dependent or environment-dependent. The evaluation results show that such strategy works effectively to enhance the physical robustness of AEs.

\subsection{Targeting Sensor Processing Module}
After passing through transmission channel, the adversarial audio will be further fed into sensor processing module, which is inside the microphones used by IVC devices.
The adversary can exploit the hardware vulnerabilities inside the sensors, such as non-linearity, etc., to launch the attack against the IVC devices.

In~\cite{46,47}, the authors identify that the headphone cable of cell phones can be targeted as an antenna, which will capture electromagnetic wave and the sensors will decode voice commands from the wave due to electromagnetic interference effect. By exploiting such vulnerability, the authors demonstrate that an adversary can easily command the victim's cell phone to turn on Wifi or Bluetooth, open a malicious website, etc., without victim's awareness. Light Commands~\cite{48} novelly leverages photo-acoustic effect to craft voice commands into laser light, which will be unintentionally captured and decoded by the microphones equipped in smart home devices such as Google Home console. Furthermore, Light Commands also demonstrates the capability of quite long distance attack, up to 110 meters.


In~\cite{38,39,40,41,42,43,44,45}, the authors exploit the non-linearity vulnerability of the microphones in IVC devices. In particular, the adversary can craft malicious voice commands at inaudible frequency by exploiting non-linearity in the demodulation of current microphone technology. IVC devices is able to decode such ``hidden'' command, which is totally imperceptible to human. Therefore, the adversary can control the target devices using such inaudible signals without victim's awareness.

\subsection{Targeting Signal Pre-processing Module}
Signal pre-processing module involves the conversion of the audio signal into digital space from physical space. The attacker can carefully modify or modulate the adversarial audio, making the digital output after the signal pre-processing module be same or similar as that of their desired phrase. 
Many researches~\cite{7,18,19,29} fall into this category.

Cocaine Noodles~\cite{19} and Hidden Voice Command~\cite{7}\footnote{Since~\cite{7} is the follow-up of~\cite{19}, Table \ref{Attack} only focuses on the white-box attack part in~\cite{7}.} are the first to explore the perceptional difference between human beings and machines when recognizing audio information. Specifically, they find that many popular speech recognition systems utilize MFCC to extract acoustic features for analysis and decoding. Therefore, they propose to mangle a malicious voice command signal, so that it can preserve enough MFCC acoustic features for the target speech recognition systems to recognize their desired command, but remain difficult for human being to interpret. They demonstrate their attacks against both white box (CMU Sphinx) and black box (Google Now) speech recognition systems, and the human perception study shows the adversarial audio is almost uninterpretable to human being.

Followed by~\cite{7,19}, the authors in~\cite{18,29} propose four classes of perturbations, which can be used to produce unnoticeable attack audio samples, based on the fact that the original audio and the revised attack audio (with perturbations) share similar speech feature vectors after being transformed by acoustic feature extraction module. Furthermore, they argue that different speech recognition systems usually share similar signal processing and feature extraction algorithms, thus their attack is model-agnostic and can be used to attack multiple black box models such as Google Speech API, IBM speech to text API, etc.

\subsection{Targeting Acoustic Model}
The acoustic features extracted by the feature extraction module will be conducted model-based prediction in ASR systems. Similar as attacks targeting pre-processing module, an adversary can generate adversarial examples to make the output results similar with those of the desired phrase.

\subsubsection{Gradient Based Approach}
A large amount of existing works~\cite{2,3,4,6,7,8,9,12,13,17, 21, 20,30,27,28, 32,33,34,36} utilize gradient descent based adversarial example generation approaches, a mathematical optimization strategy to find local minimum for a target objective loss function. Thus, the design of objective loss function plays an essential role for gradient based approaches to produce an effective adversarial example. Existing researches have validated the effectiveness of cross-entropy loss~\cite{3,12}, Connectionist Temporal Classication (CTC) loss~\cite{2,6,11,17,34,36,20}, Probability Density Function (PDF) id sequence loss~\cite{4,13} and Houdini loss~\cite{8}.

Besides causing the desired misclassificaiton on the target ASR system, another important aspect to evaluate an adversarial example is its imperceptibility to human being. Previous researches~\cite{3,11,12,27} utilize psychoacoustic hiding approach to further enhance the imperceptibility of the generated audio adversarial example. Specifically, psychoacoustic hiding approach is based on frequency masking effect in signal processing domain, i.e., the phenomenon that a louder signal can make other signals at nearby frequencies imperceptible. Therefore, the attacker can carefully design a masking threshold in the frequency domain in the audio signals, which makes other signals that fall under the threshold imperceptible to human being.

\subsubsection{Genetic Approach}
One limitation of the gradient descent based approach is that the model architecture and parameters must be known to compute the derivative of each network layer. 
However, this assumption may not be true in the scenario that the attacker intends to attack a black box model. To overcome this issue, previous researches~\cite{1,2,5,16,22,35,37} propose gradient-free approaches, mostly based on genetic algorithm. For example, the authors in~\cite{1}  propose evolutionary multi-objective optimization approach to attack two white-box ASR systems in both un-targeted and targeted settings. In~\cite{2}, the authors apply the particle swarm optimization method, an enhanced optimization genetic algorithm, to attack black box ASR models. In \cite{16}, the authors combine genetic algorithm with gradient estimate approach using CTC-loss to address the black box challenge.

\subsubsection{Universal Adversarial Perturbations}
Although researchers have developed and demonstrated many effective audio adversarial attacks, most of them are input-dependent, which means the perturbation generation strategy is based on a pre-determined, specific  input audio. This limitation makes most of existing audio adversarial attacks less effective in the scenario with streaming audio as input, since it is not likely for the attacker to solve input-dependent optimization problem in a timely manner. To address this issue, existing works~\cite{35,17,114} propose input-agnostic universal perturbation generation approaches. Intuitively, they demonstrate there exist adversarial perturbations, which can be added to any input audio signal and cause the synthesized audio misclassified by ASR systems. 

Besides input-agnostic adversarial attack, another type of universal attack is model-agnostic attack, which means the same adversarial example can fool different ASR systems. Such attack is also referred to as \textit{Transferability} attack, which will be discussed in Section~\ref{sec:transfer}. Specifically, only a few existing works~\cite{9} study the existence of model-agnostic audio adversarial examples.

\subsection{Targeting Language Model and NLP Model}
In comparison with attacks against previous modules, there are only a few existing works~\cite{23,25,26,27} targeting language model and NLP model. We believe this is because most language model and NLP model are non-differentiable, so that major gradient descent based strategies are difficult to be applied. Existing studies mostly focus on the misinterpretation attacks towards third-part applications such as Alexa Skills and demonstrate that those skills can be exploited by phrases with similar pronunciation. For example, Kumar et
al. \cite{24} present an empirical analysis of the interpretation
errors on Amazon Alexa, and demonstrate the adversary can
launch a new type of skill squatting attack. Zhang et al. \cite{25}
report a similar attack, which utilizes a malicious skill with
similarly pronounced name to impersonate a benign skill.
Zhang et al. \cite{26} develop a linguistic-guided fuzzing tool
in an attempt to systematically study such attacks.

\section{Systematization of Defense}
\label{sec:defense}
In  this  section,  we  present  the  systematization  of existing  defense mechanisms against attacks for ASR based on our modularized approach. In Table~\ref{Defense}, we elaborate the existing literature based on defense methods discussed in Section~\ref{sec:defense}. Different than attacks, which mainly exploit the vulnerabilities of one module, existing defense can either enhance one module itself (i.e., adversarial training for acrostic model) or add another defense mechanism between modules (i.e., input transformation between sensor module and signal pre-processing). Therefore, our systematic study of defenses works is organised according to the modules where these defense works are applied, i.e., ahead of sensor module, between sensor module and ASR system, as well as inside ASR system.

\begin{table*}[ht]
\centering
\small
\caption{Defense Taxonomy.}
\label{Defense}
\begin{tabular}{m{8.5cm}<{\centering}|m{4cm}<{\centering}|m{1.7cm}<{\centering}|m{1.8cm}<{\centering}}
\Xhline{3\arrayrulewidth}
\textbf{Defense} & \textbf{Target} &\textbf{Generality}  & \textbf{Knowledge} \\ \hline
\cite{66,67,68,69,70,71,72,73,74,78} & Before Sensor & Universal & None \\ \hline
\cite{49,50,51,52,53,54,55,56,57,59,60,75,76,77,79} & Between Sensor and ASR & Specific & Partial \\ \hline
\cite{33,61,64,65,80} & Inside ASR & Specific & Full \\ \hline
\Xhline{3\arrayrulewidth}
\end{tabular}
\end{table*}

\subsection{Ahead of Sensor Module}
Defense mechanism can be deployed ahead of sensor module to detect audio attacks. Specifically, such defense mechanisms can help detect physical audio attacks since digital attacks start after the sensor module. 
In~\cite{66,67,68,69,70,71,72,73,74,78}, the authors propose the audio source identification strategy to check if the captured audio is from real human or a hardware speaker. These defense solutions are promised to be effective towards audio adversarial attack since until now, all adversarial examples for the practical world attack must be played by speakers. However, one main limitation of such defense works is that the detection distance is too short. 

\subsection{Between Sensor Module and ASR system}
Before the audio signal captured by the hardware sensor is fed into ASR system, defense solutions can also be deployed to mitigate audio attacks. Previous works adopt several audio transformation methods (i.e., audio down-sampling~\cite{51}, perturbation cancellation~\cite{51,52,57,59}, adding distortion~\cite{50,53,56}, signal smoothing~\cite{57}, audio compression~\cite{49,55,60}) to destruct the adversarial perturbation (if any) to protect ASR systems. Other works apply an extra detection network~\cite{54,62,76,77,79} or multi-model detection mechanism~\cite{75}. However, those defense solutions may suffer from adaptive attack, if the adversary learns the existence of such protection.

\subsection{Inside ASR system}
Defense mechanisms can also be deployed inside ASR system, usually by adopting \textit{adversarial training} method. The idea of adversarial training is first proposed in~\cite{118} to defend image adversarial attacks, by including adversarial examples during the model training stage to improve the model's robustness towards adversarial attack. Previous works~\cite{33,61,65,80} against audio adversarial attacks also apply adversarial training strategy. Notwithstanding, we believe such solution has some limitations: i) it typically decreases the recognition accuracy of the original model and ii) it cannot effectively defend attacks targeting sensor module and pre-processing module. Besides adversarial training method, \cite{64} leverages the prediction activation inconsistency to detect the audio adversarial attack against a CNN audio recognizer, by integrating a self-verification stage inside CNN. However, such defense method needs instrumentation inside the specific CNN model.

\section{Comparison with Image Attacks}
\label{sec:comparewithimageattack}
In parallel with the security analyses on speech recognition, there is a line of research on adversarial learning for image recognition systems (IRS)~\cite{109,116,118,121,128,130}, which has been carried out earlier and therefore gained more attention. These two types of systems are different in signals they receive and processing mechanisms they run (discussed later), but 
all built on top of popular deep learning technologies and can all be considered as perceptual systems. Therefore, they are similar in their workflows, which prior research has leveraged to use attack strategies from the image realm to exploit audio systems. For example, Devil's Whisper~\cite{14} applies the momentum algorithm~\cite{127} to further enhance the transferability of crafted audio adversarial examples. However, in general, direct application of image-based attacks to speech recognition is likely to encounter unique challenges, which have not been studied before, to the best of our knowledge. In our research, we performed first such a systematic comparison study, which leads to a set of key take-aways that help better understand the challenges and guide future research on audio security. 

As mentioned earlier, our modularized system workflow includes two modules: \textit{acoustic model and language model}, right after the audio pre-processing step. By comparison, a typical image recognition system is simpler, with only a recognition neural network. During the rest of the paper, we use \textit{DNN based prediction} to refer to both the DNN module of the image system and acoustic and language models in the ASR system, for the simplicity of presentation. 



\subsection{Transmission Channel}
\label{sec:transmission with image}

Similar to the modularized speech recognition system workflow, image signals also pass through a transmission channel before captured by sensors (i.e., camera or the LiDAR system).  Like the noisy voice channel, image transmission is affected by environmental noise that reduces the success rate of an adversarial attack. Previous works~\cite{150,151,15} have proposed many effective ways to enhance the robustness of the attack, allowing adversarial examples to remain effective under various real world conditions. However, application of these techniques to elevate the robustness of audio attacks needs to consider the differences between the image and audio transmission channels. In this part, we deliberate these differences and discuss how to better understand practicability problems for audio attacks.


\vspace{1pt}\noindent\textbf{Practicability challenges}. We look into the challenges posed by physical conditions for both image and audio attacks:

\underline{\textit{Signal Reproduction Error:} } For both image and audio attacks, the adversarial sample needs to be first presented by the source (speaker for audio attack and  screens or printers for image attack), which might introduce a reproduction error.

For the physical image attack, all perturbations are required to be valid colors. Previous work~\cite{110} shows that reproduction devices like printers generate reproduction errors that reduce the success rate of a physical attack.

Reproduction errors are happen in an audio attack. As mentioned earlier~\cite{7}, performing gradient descent on audio samples often brings in large spikes to audio signals, which are extremely hard for the dedicated speaker to accurately reproduce. Therefore, to successfully reproduce the computed perturbations in a physical environment, all perturbation values should be feasible for speakers to generate and play over the air. However, to the best of our knowledge, there is no systematic research on how such signal reproduction error undermines the efficacy of the audio adversarial attack. 

\begin{tcolorbox}
\textbf{Our Take-away 1.} \textit{Like image adversarial attacks, physical audio adversarial attacks can also be affected by the signal reproduction error. A systematic research is expected to understand how this issue will impact the practicability in audio domain. }

\end{tcolorbox}

\underline{\textit{Transmission Loss:}} After the adversarial audio is played by the speaker, it would then be transmitted over the open air. During this process, the audio will suffer transmission loss caused by transformation effects like multi-path effect~\cite{111}, echo, reverberation and strength reduction, which may undermine the effectiveness of an adversarial attack. 

For the image attack, however, the loss introduced by the transmission process tends to be much smaller, which can often be ignored.


\underline{\textit{Environmental Conditions:}} In addition to the transmission loss, the adversarial audio will also be affected by other environmental conditions like ambient noise in the background, which could weaken the adversarial features in the audio. In an image attack, the adversarial image or object can be impacted by viewpoint shifts, distance and angle between it and the camera, lighting/weather conditions and others. Therefore, any perturbation must be able to survive such transformations. 

\vspace{2pt}\noindent\textbf{Existing Approaches}. As mentioned earlier (Section~\ref{sec:attack}), prior research~\cite{4,7,10,11,12,13,14,21,27,28,29,30} studies how to overcome the physical world challenges for audio adversarial attacks. Overall, the methods proposed by existing work can be categorized as either robustness enhancement or expectations over transformation (EOT). Likewise, there is a line of research~\cite{150,151,152} in the image field that also utilizes expectations over transformation (EOT) to aim at modeling the distributions of synthetic or physical transformations on the original image, so as to improve the robustness of the adversarial attack under physical conditions. Specifically, the EOT method needs to deal with different conditions. For example, in the audio domain, the attacker focuses more on how to model the transformations under the ambient noise, multi path effect and signal attenuation~\cite{21}, while image attacks need to model the transformations under distance/angle from the camera, and lighting. We believe that EOT would be a promising approach to help adversarial attack survive various physical conditions in the real world and the future physical audio adversarial attack may still consider to use this strategy to enhance its physical robustness. 

\begin{tcolorbox}
\textbf{Our Take-away 2.} \textit{Although audio and image attacks face different physical challenges, Expectations over Transformations (EOT) appear to be a promising solution that enables the adversarial attack to survive various physical conditions in the real world. }
\end{tcolorbox}


Compared with the previous work on the physical image attack, the research on the physical audio attack should consider more diverse physical-world situations. A prior study on the image attack demonstrates the feasibility in misleading the camera system using the adversarial road sign under inherently unconstrained environment (i.e., the open street), while existing physical audio attacks~\cite{12,14,21} usually limit the attack scenario to a relatively stable physical environment such as room or office with little variation in the ambient noise, multi path effect, signal attenuation and etc. Although this is a reasonable scenario for the attack on smart home devices, future audio adversarial attacks could also happen in a car (on car-based audio systems)~\cite{13} or the open-air environment like street (for devices employed outside the house), which could actually have more serious consequences. Unfortunately, the widely used Room Impulse Response modeling presented in the previous research~\cite{12,14} may not be effective under these conditions. Further research is expected to understand realistic security threats to ASR systems under the diverse environments. 



\begin{tcolorbox}
\textbf{Our Take-away 3.} \textit{Future audio physical attacks need to consider inherently unconstrained environments like inside the car or open street, instead of stable physical conditions like in a room or office. }
\end{tcolorbox}

\subsection{Sensor Processing}

In parallel with the sensor processing module introduced in Section~\ref{sec:methodology}, hardware based sensors like camera are adopted to detect and capture image information. Likewise, previous work also reports sensor-level attacks~\cite{136,137} on image recognition system. Such attacks have been covered by a SoK paper on sensor security~\cite{112}.

\subsection{Pre-processing}
In both image and audio recognition systems, raw signals are passed through the pre-processing module after they are captured by sensors. For an image, pre-processing operations remove irrelevant information and identify useful image features. Similar to the ASR system, the IRS system also performs denoising and transformations on digital signals. For this purpose, median filter~\cite{204}, mean filter~\cite{205} and Gaussian filter~\cite{206} are typically used to remove the impulse noise from sudden disturbances, additive white Gaussian noise, etc. Then, image geometric transformation like translation, rotation, scaling, mirroring, etc., is performed on the visual signals to correct the errors introduced by sensors or measurement. Furthermore, the pre-processing module runs down-scaling algorithms to handle those images that do not match the model’s input size. For example, the LeNet-5~\cite{lecun2015lenet} architecture needs to fix the image size to 32x32 so any input is scaled to this dimension. Finally, useful information in the image signal needs to be enhanced: e.g., the high pass filter can be used to emphasize the edges/details thereby making a blurry object clear, while the low pass filter can be leveraged to denoise thereby smoothing the entire image.

For audio signals, similar operations are also performed by popular systems. For instance, the Kaldi Aspire~\cite{kaldi} model needs an 8k sample rate for the input audio, so the audio signals with a higher sample rate will be down-sampled for further processing. This module, however, is subject to the \textit{down-scaling attack} in the image domain, as demonstrated by Xiao et al.~\cite{154} and Erwin et al.~\cite{155}. Specifically, the attack manipulates an image to get different visual results so as to change the semantic meanings when scaled to a specific dimension. Even worse is that unlike the adversarial examples attack, the image scaling attack is model agnostic since existing learning algorithms typically require fixed input dimensions. As a result, such an attack could potentially affect a wide range of image applications and systems.

We believe it feasible that the strategy employed by the image scaling attack could be applied to manipulate audio signals, as the similar audio down-sampling operation is also performed during audio pre-processing. Actually, Yuan et al.~\cite{4} and Chen et al.~\cite{13} propose that audio down-sampling could serve as a defense mechanism against their audio adversarial attacks, since the crafted adversarial perturbations could be mitigated by the operation. These studies show that audio down-sampling could cause information loss from the original signals, which conforms to the assumption underlying the image scaling attack. However, to the best of out knowledge, there is no report that leverages this strategy to attack ASR. Therefore, we consider this to be an interesting research direction, since this type of attack is model-independent and could potentially compromise various speech recognition systems and devices.

\begin{tcolorbox}
\textbf{Our Take-away 4.} \textit{Inspired by the image scaling attack, audio signal loss caused by down-sampling operations may potentially lead to “audio cognition contradiction” between humans and machines, which would be a powerful new approach for unnoticeable audio attacks.}
\end{tcolorbox}


\subsection{DNN based Prediction}
One of the most important components of both ASR and IRS systems is machine-learning-based prediction, which mainly computes the probability score for each potential class output on given input features. Similar to the attacks targeting machine learning models of the ASR system, adversarial attacks can also happen on the IRS system, which has been widely studied~\cite{109,116,118,121,128,130}. By comparison, the research on the adversarial attacks on the ASR system is still limited, due to the complexity of the ASR system and the challenges brought by audio signals: (i) One traditional ASR system usually contains more components than an IRS system, for instance, time-frequency domain transformation, MFCC feature extraction, language model and NLP model for language information decoding. Although those additional components introduce more opportunities and interfaces to attack (i.e., ~\cite{25,26,27} targeting language and NLP model), we believe that they could also make it more difficult to perform the end-to-end audio adversarial attack. It is mainly because the undesired interactions among those components could potentially undermine the adversarial perturbation information. (ii) Audio signals fed into ASR systems tend to have a high sample rate (i.e., 8k sample rate for Kaldi Aspire model and 16k sample rate for DeepSpeech model), while images usually have only hundreds of pixels information (i.e.,28x28 pixel size for LeNet~\cite{lecun2015lenet} model input). Thus, we believe that such a huge difference for the input dimensions between the image signal and audio signal may significantly raise computational cost when porting previous methods from the adversarial image attack to audio domain.

Here, we present our findings and take-aways on how to improve the efficacy of audio attacks on machine learning models, by comparing with existing image adversarial attacks.

\vspace{2pt}\noindent\textbf{Which components of the input should be crafted?} Prior audio adversarial attacks~\cite{4,6} add generated perturbations to the whole piece of the original signal. However, one may ask: \textit{is it necessary to modify the whole signal to produce an effective adversarial example?} Ideally, the perturbation size should be limited to avoid being noticed by a human. Unfortunately, to the best of our knowledge, no previous research in the adversarial audio attack community reports the findings on this subject. In the image domain, JSMA approach~\cite{109} and One-pixel attack~\cite{113} prove that by only changing a little bit of the whole pixels~\cite{109} or just only one pixel~\cite{113}, the adversary could still effectively fool the target image classifiers. Specifically, in the JSMA approach, Papernot et al. propose a method to generate image adversarial examples using forward derivatives of the target neural network. Intuitively, the authors consider that not all input domain regions play the important role in generating effective adversarial examples. So, they propose to compute the forward derivatives of a DNN to construct the adversarial saliency map. This could indicate which input feature components lead to significant changes regarding the neural network output results. Instead of changing the whole input feature spaces, they could only modify a small portion of the input features to cause the target model's misclassifications. In the One-pixel attack, Su et al. propose a more extreme approach: only modifying one pixel of the image. Intuitively, they adopt a differential evolution algorithm to find the optimal solutions under the assumptions. These prior studies show that \textit{not all regions from the input domain are contributing to adversarial example discovery.} We believe that this observation is still true for audio-based attacks since the DNN part of the image and audio systems share similar weakness. Thus, such idea may be leveraged for audio attacks to reduce the perturbation size, thus rendering the adversarial samples less perceivable.

\begin{tcolorbox}
\textbf{Our Take-away 5.} \textit{Previous research in the image field demonstrates that not all regions from the input domain contribute to an adversarial sample's successful attack. So we need to consider how to efficiently select the components of the sample that would be meaningful for successful attacks. However, computation cost needs to be effectively controlled to make the approach feasible for an audio attack. }
\end{tcolorbox}

However, applying this strategy to an audio attack is non-trivial since the audio signal itself tends to be more complicated than the image signal. For example, the LeNet~\cite{lecun2015lenet} model used in JSMA approach~\cite{109} takes the input of black and white images (28x28 pixels) which can be further processed as 784 features. In contrast, for a normal speech recognition model like Kaldi Aspire model~\cite{kaldi}, it takes audio input of the 8k sample rate, leading to 8000 sample points for one-second audio (the regular length for many target phrases). Also, the DNN output of Kaldi Aspire model gives 8629 probability scores based on all potential speech pronunciations. In comparison, LeNet model outputs ten scores for each possible class (Arabic number 0 to 9 for that handwriting digits recognition model). As stated in~\cite{109}, the JSMA method runs slowly due to its significant computation overhead. So we believe that one needs to propose an efficient algorithm to dramatically reduce the computation cost of this approach to enable an audio attack.


\vspace{2pt}\noindent\textbf{Objective function establishment.} Intuitively, for most white-box based attacks, adversarial examples are typically crafted by performing a gradient descent algorithm with respect to the input on an 
\textit{objective function} designed to be minimized to make the input adversarial towards the target model. Even for some black-box adversarial attacks which adopt genetic algorithm~\cite{1,5,37} or gradient-free optimization approach~\cite{115}, one core of the attack is still the objective function (usually as fitness score for the genetic algorithm), which performs as the modification direction indicator for generating adversarial examples. As objective function can significantly affect the sample's modification direction, we believe that how to select or establish the objective function of the attack would be one interesting question that may lead to effective and transferable attacks, which is also discussed in~\cite{116}. Previous works on audio adversarial attack has explored various objective functions to be applied, for instance, CTC-Loss~\cite{2,16,20}, cross-entropy Loss~\cite{3,6,12}, MFCC vectors loss~\cite{7,19}, pdf-id (probability density function identifiers) loss~\cite{4,13} and Houdini Loss~\cite{8}. In image adversarial attack area, Carlini et al.~\cite{116} conduct a thorough study towards the attack results under different objective functions. In~\cite{117}, Sabour et al. propose to concentrate on the internal layers of DNN representations instead of focusing on the output label of the DNN model. Compared with previous work in the image field, we consider there lacks such depth study for selecting the objective function, as the objective function may significantly affect the effectiveness and transferability of the generated adversarial examples.

\begin{tcolorbox}
\textbf{Our Take-away 6.} \textit{Since the objective function may significantly affect the effectiveness and transferability of the generated audio adversarial examples, it is essential to understand the implications of the function when protecting ASR systems.}
\end{tcolorbox}

\vspace{2pt}\noindent\textbf{Other insights.} Besides the above-mentioned insights, there are also some thought-provoking questions to be discussed left for future studies. First, compared with image adversarial attacks which normally target one single label (i.e., dog or panda), audio adversarial attacks usually need the target model to produce a phrase or sentence (i.e., "Echo, open the door"). This introduces more challenges when porting the attacks from the image domain to the audio domain. Second, the audio recognition system tends to be more complicated than the image recognition system, e.g., having time-frequency domain transformation, MFCC extraction, and language model, which increase the difficulties of the attacks. Finally, unlike the L-p norm metrics used in the image area, there lacks adequate metrics to measure human perception on the added audio perturbations. So future attacks may need to take this into considerations.


\section{Comparison with Defenses in Image Domain}
\label{sec:comparewithimagedefense}
Countermeasures for attacks (mainly adversarial attacks) towards IRS system have been well studied in past years. Similar with the previous section, this section is still organised according to the modularized workflow. We present what we can learn from the defense in the domain of images.

\subsection{Before the Sensor Module}
While liveness detection~\cite{66,67} could distinguish whether the upcoming audio signal is adversarial or not, image signals cannot be detected in this way, since adversarial image does not contain enough clue to distinguish machines or humans. Thus, to the best of our knowledge, in the field of image processing, no defense has been deployed before the sensor module in IRS system.

\subsection{Between Sensor Module and IRS system}
Before the signal is fed into a target IRS system, the defense strategies of input transformation/reconstruction can be deployed. It is similar to the defenses in the domain of audios, mitigating the image adversarial attacks. Such works include~\cite{119} and~\cite{120}, which are demonstrated to be effective.


\subsection{Inside IRS system}
Similar to the adversarial training in the audio domain, such a method is also widely adopted in the image domain to enhance the robustness of neural networks~\cite{121,122,123}. However, two other popular defenses in the image domain are not adopted in the audio domain: gradient masking~\cite{124} and defensive distillation~\cite{125}.
Gradient masking aims to train another derivable model by penalizing the degree of variation. So the imperceptible perturbation from the adversarial sample will not be likely to change the label unexpectedly. 
Defensive distillation is to transfer a DNN from a large neural network to a smaller one. It changes the way of the model to handle inputs, potentially defending against the AEs.
To the best of our knowledge, yet no previous work explores the two defenses in the domain of audio. Maybe it is because of the neural network's complexity to handle audio inputs, which lets the training process bear a high cost. Hence, before studying the two methods' feasibility, reducing the training cost should be done first.


\begin{tcolorbox}
\textbf{Our Take-away 7.} \textit{Network distillation and gradient masking are not well studied in the audio domain. Before adopting the two defenses, researchers should handle the problem of high training cost.}
\end{tcolorbox}

\section{Towards the Transferability}
\label{sec:transfer}
Adversarial examples (AEs) could transfer across different models: AEs generated on one specific model could fool other unseen models. As a consequence, the attacker could leverage it to attack the target black-box models, even though the knowledge of the black-box models remains unknown. The nature of adversarial transferability is first proposed and investigated in~\cite{118}, and has gained lots of interest during recent years. The successful transferable adversarial attacks have been reported in many previous works~\cite{118,127,128,129,130,131,132,133,134} targeting IRS systems. However, only limited studies~\cite{13,18,19} put attention on transferable audio adversarial attacks. This section will discuss how future work could address the challenge of transferability for audio attacks, from the knowledge of existing image transferable attacks. Particularly, we show that transferable audio attacks based on gradient method is feasible, which contradicts the consequence from~\cite{126}.


\subsection{Image-domain Inspired Methods on Transferability.}
Intuitively, the attacker could borrow the approaches from existing image transferable attacks to make audio transferable adversarial attacks.

First, the attacker could train \textit{Surrogate Model} to approximate the target model, and then craft adversarial samples using the surrogate model. Theoretically, such adversarial samples are more likely to transfer to the target model. Previous works~\cite{129,130} demonstrated this method in the image field. However, training the surrogate model in the audio domain is generally more complicated than the image domain. The size of training data for ASR models is usually much larger than the data for image models, making it difficult to approximate the target model in the audio domain. Intuitively, in~\cite{129} and~\cite{130}, the authors showed that they could train the full simulated model on a given training dataset after hundreds of queries towards the target black-box model. In~\cite{13}, the authors adopted a similar idea to approximate the target audio black-box model. However, the results showed that the generated adversarial samples by the surrogate model had a low transferability towards the target model. The authors proposed to further enhance the surrogate model with a large open-source ASR, which helps to address this problem. Therefore, we believe that training a single model to approximate the target audio black-box model under limited queries would be an interesting future research direction. 

Second, the attacker could use \textit{Ensemble Training} strategy, which was proposed in~\cite{128}. Apparently, if an adversarial sample can attack multiple models,  it has a higher possibility to attack other unknown models. To the best of our knowledge, there is no proposed work using this idea to attack. 

Third, previous works~\cite{127,131,134,135} have reported empirical findings on how to optimize the AE generating algorithm to enhance the transferability. Among these researches, many of them~\cite{127,131,134} aim to optimize gradient direction to improve the transferability of AEs. In~\cite{127}, the authors introduced momentum during every iteration procedure when crafting adversarial samples, which could stabilize update directions and escape from poor local maxima, further enhancing the transferability. In~\cite{133}, the authors show that they can improve adversarial transferability by increasing the perturbation on a pre-specific layer of the model. In~\cite{134}, the authors propose Curls-Whey black-box attack to combine gradient ascent and descent directions together, hoping to escape overfitting direction and further find more possible solutions to pass the black-box decision boundaries.

Last, besides gradient direction optimization, the attacker could also improve the adversarial sample's robustness for multiple models. In~\cite{135}, the authors applied random transformations to the input images at each iteration in the attack process to improve the transferability. In~\cite{133}, the authors proposed variance-reduced iterative gradient sign method (vr-IGSM), which aims to remove the local fluctuation of the gradient and make adversarial samples robust to Gaussian perturbation in the meanwhile, so to enhance the samples' transferability.

We consider that the above ideas could also be ported to audio domain attacks. In this process, the attacker should notice the enlarged size of perturbation and the impact on human perception.

\subsection{Audio Tranferability Experiments}
Considering transferability of gradient-based approaches towards ASR system is still a large space for research, we conduct extensive experiments to study if adversarial attacks based on gradient-based approaches could transfer across ASR models or not. Contrary to the opinion stated in~\cite{126}, we believe that adversarial features' robustness may play a more critical role than model similarity (including hyper-parameters, architecture, training set) concerning transferability. A recent study CommanderSong~\cite{4} uses a random noise model to improve the robustness of the AEs, letting it transfer from Kaldi ASpIRE Chain Model to iFlytek. So, we consider that in~\cite{126}, the adversarial features might be too specific or overfitting to the DeepSpeech model, thus losing transferability towards other models. We adopt the same gradient-based approach chosen in~\cite{126} to perturb audio samples towards the DeepSpeech model to validate our assumption. Then we use the methods from~\cite{14} to further enhance the robustness of adversarial features. We aim to check if the AEs crafted this way could transfer to other models.




\begin{table*}[h]
\centering
\small
\caption{Transferablity of the ``robust audio'' based on DeepSpeech 0.1.0 to other platforms.}
\label{Transferablityofrobustae}
\begin{tabular}{ccclcccl}
\Xhline{3\arrayrulewidth}
\multicolumn{1}{c|}{\multirow{2}{*}{\textbf{Target phrase}}} & \multicolumn{1}{c|}{\multirow{2}{*}{\textbf{\begin{tabular}[c]{@{}c@{}}Audio \\ length\end{tabular}}}} & \multicolumn{1}{c|}{\multirow{2}{*}{\textbf{\begin{tabular}[c]{@{}c@{}}Success number\\ on DS-0.1.0\end{tabular}}}} & \multicolumn{4}{c|}{\textbf{Transfer AEs number on other platforms}}& \multicolumn{1}{c}{\multirow{2}{*}{\textbf{SNR (dB)}}} \\ \cline{4-7}
\multicolumn{1}{c|}{} & \multicolumn{1}{c|}{} & \multicolumn{1}{c|}{} & \multicolumn{1}{c|}{\textbf{DS-0.2.0}} & \multicolumn{1}{c|}{\textbf{DS-0.3.0}} & \multicolumn{1}{c|}{\textbf{Microsoft API}} & \multicolumn{1}{c|}{\textbf{Microsoft web}} & \multicolumn{1}{c}{} \\ \hline
\multicolumn{1}{c|}{\textbf{where is my car}} & \multicolumn{1}{c|}{2s} & \multicolumn{1}{c|}{141} & \multicolumn{1}{c|}{140/141} & \multicolumn{1}{c|}{140/141} & \multicolumn{1}{c|}{120/141} & \multicolumn{1}{c|}{90/141} & \multicolumn{1}{c}{4.89$\sim$12.15} \\ \hline
\multicolumn{1}{c|}{\textbf{how is the weather}} & \multicolumn{1}{c|}{2s} & \multicolumn{1}{c|}{78} & \multicolumn{1}{c|}{13/78} & \multicolumn{1}{c|}{12/78} & \multicolumn{1}{c|}{75/78} & \multicolumn{1}{c|}{71/78} & \multicolumn{1}{c}{5.11$\sim$6.16} \\ \hline
\multicolumn{1}{c|}{\textbf{open the front door}} & \multicolumn{1}{c|}{2s} & \multicolumn{1}{c|}{59} & \multicolumn{1}{c|}{10/59} & \multicolumn{1}{c|}{10/59} & \multicolumn{1}{c|}{13/59} & \multicolumn{1}{c|}{11/59} & \multicolumn{1}{c}{5.27$\sim$8.5} \\ \hline
\multicolumn{1}{c|}{\textbf{open the website}} & \multicolumn{1}{c|}{2s} & \multicolumn{1}{c|}{69} & \multicolumn{1}{c|}{0} & \multicolumn{1}{c|}{0} & \multicolumn{1}{c|}{30/69} & \multicolumn{1}{c|}{3/69} & \multicolumn{1}{c}{3.85$\sim$7.3} \\ \hline
\multicolumn{1}{c|}{\textbf{turn off the light}} & \multicolumn{1}{c|}{2s} & \multicolumn{1}{c|}{92} & \multicolumn{1}{c|}{0} & \multicolumn{1}{c|}{0} & \multicolumn{1}{c|}{87/92} & \multicolumn{1}{c|}{69/92} & \multicolumn{1}{l}{4.86$\sim$9.27} \\ \Xhline{3\arrayrulewidth}
\end{tabular}
\begin{tablenotes} 
\footnotesize
\item[1] \textbf{Note:} (1) We find it is difficult for DeepSpeech is to recognize the first word of the target phrase, which maybe because of the pre-processing module (such as voice activity detection and de-noise, etc). Therefore, we add an auxiliary word (i.e., hey) before the target phrase, and the auxiliary word is not considered for the success rate. (2) We check the top-10 transcriptions of the DeepSpeech model, and and top-5 transcriptions of Microsoft Bing Speech Service API. If the target phrase can be found at least once in the top-K  transcriptions, we take the AE can success attack the model. (3) ``DS-0.1.0'', ``DS-0.2.0'' and ``DS-0.3.0'' are short for DeepSpeech 0.1.0., DeepSpeech 0.2.0 and DeepSpeech 0.3.0, respectively. (4) Microsoft API and Micorsoft web represent Microsoft Bing Speech Service API API and the website (https://azure.microsoft.com/en-us/services/cognitive-services/speech-to-text/features). (5) We choose 10 daily command as the target phrase. For each target phrase, we use the 2-second clip of a song as the original audio. And saved 150 audios after 1500 iterations of the ``robust AE'' attack. Generally, for the 5 phrases in this Table, we can generate transferable AEs to attack Microsoft API and Microsoft web. Transfering to other versions of DeepSpeech model is a little difficult. As we only generate 150 samples for the target phrases based on an original audio. We did not generate the AEs for other 5 target phrases. For example, we just found some crafted audios (AEs) can be recognized as ``appointment to my calender'', ``to my home'', `the new version'', ``bluetooth'' and ``one one''. While the total target phrases are ``add an appointment to my calender'', ``navigate to my home'',``share the new version'', ``connect bluetooth'' and ``call nine one one''.  
\end{tablenotes} 
\vspace{-0.3cm}
\end{table*}

\vspace{2pt}\noindent\textbf{Transferablity of the AEs based on end-to-end ASR model.}
In~\cite{126}, Hadi et al. used Carlini's attack~\cite{6} and PGD attack~\cite{138} to generate more than 17000 AEs against a DeepSpeech model, but find that they cannot attack the other 9 DeepSpeech models. To further improve the transferability of our AEs, based on the gradient-based method in~\cite{6}, we then used the noise model from~\cite{14}, and set the bandwidth of the bandpass filter as 200 Hz$\sim$4000Hz\footnote{To improve the stealthy of the injected commands, the bandwidth of the bandpass filter is set as 1000 Hz$\sim$4000Hz in~\cite{14}. Choosing 200 Hz as the lower sideband helps us generate more AEs.}. Our original target model is DeepSpeech 0.1.0. We saved one AE for every ten iterations. In sum, we can get 150 AEs after 1500 iterations for each command. Then, we take those successful AEs for DeepSpeech 0.1.0 model and give them to other models, including other versions of DeepSpeech and commercial black box ASR (Microsoft API and WEB services). We choose ten commonly used phrases as the target commands and the results are shown in Table~\ref{Transferablityofrobustae}. From the results, we can see 5 over 10 commands which can transfer to other models at a moderate SNR level.

\begin{table}[h]
\centering
\small
\caption{Alignments of the recognized word ``weather''.}
\label{Alignments}
\begin{tabular}{m{2.4cm}<{\centering}|m{5.2cm}<{\centering}}
\Xhline{3\arrayrulewidth}
\textbf{Audios} &  \textbf{Alignment} \\ \hline
\textbf{TTS(famale)} &  \text{wweeaa---tthherrr-----} \\ \hline
\textbf{TTS(male)} & \text{ wweea--ttherr-----} \\ \hline
\textbf{AE1\_based\_1s} &  \text{wwea--t-h-err-----} \\  \hline
\textbf{AE2\_based\_1s} &  \text{weea-tthherr-----} \\ \hline
\textbf{AE1\_based\_2s} &  \text{w-ea---tttheerr-----} \\ \hline
\textbf{AE2\_based\_2s} &  \text{ wweaa----tthherr-----} \\
\hline
\textbf{AE1\_based\_4s} &  \text{ww-e--------a---------th-e-------rr-----} \\\hline
\textbf{AE2\_based\_4s} &  \text{w-e---------a-------tthee--------rr-----} \\
\Xhline{3\arrayrulewidth}
\end{tabular}
\end{table}

Additionally, we observed that the target phrase's alignments are significantly different for various original audio lengths during generating the samples. Table~\ref{Alignments} shows the alignments of the word ``weather’’ for different audios. It can be seen that the alignments of the AEs based on 1-second (e.g., ``AE1\_based\_1s") and 2-second (e.g., ``AE1\_based\_2s") are more close to the TTS audio than the AEs based on 4-second. The duration of the ``weather’’ is about twice of the normal pronunciation. Take an extreme example, ``w-e-a-ther’’ can be recognized as ``weather’’, while it only lasts 40 frames (about 120$ms$). Obviously, it is hard to speak ``weather’’ in 120$ms$. Correspondingly, the generated features in the 120ms length are quite different from the normal pronunciation features. Therefore, the alignment will impact the transferability of the audios with various lengths, since ASR models have different alignment ways for decoding. In our tests, we use 2-second audio, which could give us better results than other audios with different lengths. We believe that how the alignment impacts the transferability would be an interesting future direction to explore.

\vspace{2pt}\noindent\textbf{Transferablity of the AEs based on classical ASR model.} 
A recent study Devil’s Whisper~\cite{13} improves the transferability of CommanderSong attack using Momentum based Iterative Fast Gradient Method (MI-FGM) and ``local model approximation''. In our study, we used the code from Devil’s Whisper to generate AEs. For two target phrases ``open YouTube'' and ``navigate to my house'',  we successfully generated AEs which can attack both Google Speech-to-Text API and Microsoft Bing Speech Service API. Only 100 iterations were needed to generate the adversarial samples~\footnote{We set the noise model value and the clip of the delta as 6500 and 3500, respectively.}.

\subsection{Summary}
To sum up, we combine the same gradient-based method used in the study~\cite{126} and robustness improving method used in~\cite{14}, and generate adversarial samples targeting DeepSpeech v0.1.0. We demonstrate that they can also transfer to other versions of DeepSpeech models or even commercial black-box ASR systems (e.g., Microsoft API and Microsoft Web). Furthermore, we show that using the method in Devil’s Whisper, the adversarial samples targeting the Kaldi model could also fool both Google API and Microsoft API. These experiments validate that \textit{gradient-based audio adversarial samples could transfer even if the model similarity is unknown.}

\begin{tcolorbox}
\textit{1. Gradient based audio transfer attacks is feasible, even if the model similarity is unknown. \\ 
2. The alignment of words may impact the transferability.}
\end{tcolorbox}

\section{Conclusion}
In this paper, we conduct a systematic and comprehensive investigation of existing works on ASR security. Specially, we propose our modularized approach and present a taxonomy of attacks and defenses towards ASR system. Moreover, we offer a novel perspective to provide insightful suggestions for future studies on ASR security through the lens of well-studied IRS security. Last but not least, we demonstrate that a transferable audio adversarial attack is achievable even on black-box ASR systems.



{
\bibliographystyle{plain}
\footnotesize
\bibliography{bibliography.bib}
}

\end{document}